\documentclass[PRX,reprint,amsmath,amssymb,superscriptaddress,longbibliography]{revtex4-1}
\usepackage{amsmath}% Physical Review X
\usepackage{amssymb}
\usepackage{graphicx}
\usepackage{dcolumn}
\usepackage{bm}

\usepackage{color}

\begin{document}

\title{Heavy fermion quantum criticality at dilute carrier limit in CeNi$_{2-\delta}$(As$_{1-x}$P$_{x}$)$_{2}$}
\author{Jian Chen}
\affiliation{Department of Physics, Zhejiang University, Hangzhou 310027, China}
\affiliation{Department of Basic Teaching, Zhejiang University of Water Resources and Electric Power, Hangzhou 310018, China}
\author{Zhen Wang}
\affiliation{Department of Physics, Zhejiang University, Hangzhou 310027, China}
\author{Yupeng Li}
\affiliation{Department of Physics, Zhejiang University, Hangzhou 310027, China}
\author{Chunmu Feng}
\affiliation{Department of Physics, Zhejiang University, Hangzhou 310027, China}
\author{Jianhui Dai}
\affiliation{Department of Physics, Hangzhou Normal University, Hangzhou 310036, China}
\author{Guanghan Cao}
\affiliation{Department of Physics, Zhejiang University, Hangzhou
310027, China}
\author{Zhu-An Xu}
\email{zhuan@zju.edu.cn}
\affiliation{Department of Physics, Zhejiang University, Hangzhou 310027, China}
\affiliation{Zhejiang California International NanoSystems Institute, Zhejiang University, Hangzhou 310058, China}
\affiliation{Collaborative Innovation Center of Advanced Microstructures, Nanjing 210093, China}
\author{Qimiao Si}
\email{qmsi@rice.edu}
\affiliation{Department of Physics and Astronomy, Rice University, Houston, Texas 77005, USA}
\date{\today}

\begin{abstract}
We study the quantum phase transitions in the nickel pnctides,
CeNi$_{2-\delta}$(As$_{1-x}$P$_{x}$)$_{2}$ ($\delta$ $\approx$ 0.07-0.22).
This series displays the distinct heavy fermion behavior in the
rarely studied parameter regime of dilute carrier limit. We systematically investigate
the magnetization, specific heat and electrical transport down
to low temperatures. Upon increasing the P-content, the
antiferromagnetic order of the Ce-4$f$ moment is suppressed
continuously and vanishes at $x_c \sim$ 0.55. At this doping,
the temperature dependences of the specific heat and longitudinal
resistivity display non-Fermi liquid behavior. Both the
residual resistivity $\rho_0$ and the Sommerfeld coefficient
$\gamma_0$ are sharply peaked around $x_c$. When the P-content
reaches close to 100\%, we observe a clear low-temperature
crossover into the Fermi liquid regime. In contrast to what
happens in the parent compound $x$ = 0.0 as a function of pressure,
we find a surprising result that the non-Fermi liquid behavior
persists over a nonzero range of doping concentration, $x_c<x<0.9$.
In this doping range, at the lowest measured temperatures,
the temperature dependence of the specific-heat coefficient is
logarithmically divergent and that of the electrical resistivity
is linear. We discuss the properties of
CeNi$_{2-\delta}$(As$_{1-x}$P$_{x}$)$_{2}$ in comparison with those
of its 1111 counterpart, CeNi(As$_{1-x}$P$_{x}$)O.
Our results indicate a non-Fermi liquid phase in the global phase diagram of heavy
fermion metals.

\end{abstract}


\maketitle


\section{INTRODUCTION}

As a result of a continuous phase transition at zero
temperature, quantum criticality has been broadly discussed in
connection with remarkable low temperature properties such as
non-Fermi liquid (NFL) and unconventional superconductivity in a
number of strongly correlated electron systems\cite{1,2,3}.  In
Kondo lattice systems, quantum criticality is attributed to
the interplay between the Ruderman-Kittel-Kasuya-Yosida (RKKY)
interaction and the Kondo couplings. The RKKY interaction between the
localized $4f$ electrons, mediated by hybridized conduction
electrons, stabilizes the 4$f$ moments so that the system tends to
show magnetic order at low temperature. On the other hand, if the
Kondo effect is dominant, the 4$f$ moments are screened by
conduction electrons so that the Fermi liquid behavior appears
at low temperature. In the transition regime, the competition between the
two types of interactions are apparent at temperatures below the coherence
temperature, where the effect of the coupling between the local moments and
conduction electrons sets in. The ground state due to the competition
between the RKKY interaction and the Kondo effect can be generally
understood by the Doniach phase diagram\cite{Doniach}.  A magnetic
quantum critical point(QCP) may arise in this phase diagram by varying
the Kondo coupling or density of charge carriers at the Fermi
energy, although its existence is always complicated in
realistic materials. The energy scale for the Kondo effect, Kondo
temperature $T_K$, depends on the $c$-$f$ hybridization and the
density of states at the Fermi energy, and thus can be tuned by
non-thermal parameters such as external pressure, chemical doping or
applied magnetic field, etc. This allows to probe a possible QCP by
identifying the associated NFL behavior in its vicinity\cite{4,5}.

In this context, what has been little explored is the limit where
the charge carriers are very dilute. In this regime, one may be
concerned that the number of conduction electrons is insufficient
to screen all the local moments completely. This is the so-called
Nozi\`{e}res' exhaustion problem\cite{Nozieres98}.  The idea behind
this problem is that the collective Kondo screening can still take
place in the limit of dilute conduction electrons but with delayed
development of coherent paramagnetic Kondo singlet state.  This
scenario has recently been evidenced in the cerium based nickle
arsenide, CeNi$_{2-\delta}$As$_{2}$\cite{Luo2012},  where the
unconventional quantum criticality has been revealed by tuning
the physical pressure\cite{Luo2015}.  More generally, the dilute carrier concentration
can lead to new types of underlying heavy fermion state and the
associated quantum phase transitions \cite{Feng2016}.
In order to explore this physics,
it is desirable to have as large a coherence temperature as
possible. Because the coherence temperature is increased by the
isovalent P for As substitution \cite{Luo2012}, we have been
motivated to study the antiferromagnetic (AFM) quantum phase transition
by the effect of chemical pressure in
CeNi$_{2-\delta}$(As$_{1-x}$P$_{x}$)$_{2}$.

CeNi$_{2-\delta}$As$_{2}$, with $\delta$ $\approx$ 0.28, crystallizing
in the well-known ThCr$_{2}$Si$_{2}$-type structure ($I4/mmm$, No.139),
was found to be a Kondo compound, with AFM transition
temperature $T_{N}$ = 4.8 K\cite{Luo2012}.  When the magnetic field
was applied along the $c$-axis, the well localized moments of Ce
$4f$ electrons undergo a spin-flop metamagnetic transition (MMT),
from an AFM to a polarized paramagnetic (PM) ground state. Moderately
strong electronic correlation is indicated by an enhanced electronic
Sommerfeld coefficient $\gamma_{0}$ $\approx$ 65 mJ/mol$\cdot$K$^{2}$
and a Kondo temperature $T_{K}$ $\sim$ 4K, rendering
CeNi$_{2-\delta}$As$_{2}$ as a new system for unraveling the competition
between the RKKY interaction and Kondo effect. In fact, two possible
QCPs in CeNi$_{2-\delta}$As$_{2}$ were reported previously:
\cite{Luo2012,Luo2015} one is induced by hydrostatic pressure at
$p_{c}$ = 2.7 GPa and the other by magnetic field at $B_{c}$ = 2.8 T,
the latter
likely being
a first-order quantum critical end point. In
particular, due to the possible Ni vacancies in this compound under
the ambient pressure, the density of charge carriers is expected to
be small.

On the other hand, the isostructural compound CeNi$_{2-\delta}$P$_{2}$
was found to be a nonmagnetic, intermediate valence Kondo lattice metal \cite{CeNi2P2}.
CeNi$_{2-\delta}$As$_{2}$ and CeNi$_{2-\delta}$P$_{2}$ are thus
located at the opposite sides of the QCP in the Doniach phase diagram.
Therefore, it is natural to consider the P substitution for As in the
Ni-As layer of CeNi$_{2-\delta}$As$_{2}$. As P has a smaller ionic
radius than As, this isovalent substitution is expected to introduce
chemical pressure and could shift the system from the AFM to the
non-magnetic ground states. If the Ni vacancies do not change very
much under the As$/$P substitution, we anticipate that
CeNi$_{2-\delta}$(As$_{1-x}$P$_{x}$)$_{2}$ should be a suitable system
to investigate whether there is a QCP in a Kondo lattice with a low
carrier density.

In this paper, we report a comprehensive study on the Kondo
compounds CeNi$_{2-\delta}$(As$_{1-x}$P$_{x}$)$_{2}$ ($\delta$
$\approx$ 0.07-0.22) using chemical doping concentration $x$ as
a tuning parameter. A $T$-$x$ phase diagram is then determined
for $0\leq x \leq 1$, and evidence for the low carrier density
is found for $0.1 \leq x\leq 0.7$. From the results of magnetic
susceptibility $\chi(T)$, specific heat coefficient $C(T)/T$ and
electrical resistivity $\rho_{xx}(T)$, we find that the AFM order
is suppressed continuously and disappears at $x_{c}$ $\approx$
0.55. Around this QCP, the NFL behavior is exhibited in $\rho_{xx}(T)$
and $C(T)/T$, which indicates a divergent effective mass. We find
the surprising
result that the NFL behavior persists over a nonzero range of
doping concentration; the Fermi liquid behavior is not recovered
until $x \geq 0.9$. In the doping range of $0.55<x<0.9$, NFL
behavior is observed at low temperatures: the electrical resistivity
is linear in temperature and the specific-heat coefficient is
logarithmically divergent. We discuss the properties of
CeNi$_{2-\delta}$(As$_{1-x}$P$_{x}$)$_{2}$ together with those of its 1111
counterpart, CeNi(As$_{1-x}$P$_{x}$)O, and draw implications for
the global phase diagram of the heavy fermion metals.

\section{Experimental methods}

Polycrystalline samples of CeNi$_{2-\delta}$(As$_{1-x}$P$_{x}$)$_{2}$
were prepared by solid state reaction method in vacuum. The raw
materials are cerium pieces (99.8$\%$), nickel powder (99.999$\%$),
and phosphorus powder (98.9$\%$) from Alfa Aesar and arsenic pieces
(99.995$\%$) from Aladdin. First, NiAs, NiP, CeAs and CeP were
pre-synthesized in a molar ratio of 1:1 in an evacuated quartz
tube at 973 K, 973 K, 1323 K and 1173 K, respectively. Second,
the initial atomic ratio of 1:1.76:2 were weighted, mixed well,
ground, pelletized and installed in Al$_2$O$_3$ crucibles. Then
the crucibles were sealed in evacuated quartz ampoules which were
sintered in vacuum at 1323 K to 1570 K for at least 2 days followed
by furnace cooling. Finally, the samples were thoroughly ground,
cold pressed and annealed in vacuum to improve the homogeneity.
Special attentions were paid to the parent compound
CeNi$_{2-\delta}$As$_{2}$ and the non-magnetic counter compound
LaNi$_{2}$As$_{2}$: the cold-pressed pellet of the precursors
was heated at 1570 K for 2 days without further annealing progress
in order to stabilize the low temperature phase with
ThCr$_2$Si$_2$-type structure according to Ref.\cite{Ghadraoui}.
It's noted that since a nickel defect dose not occur for LaNi$_{2}$As$_{2}$,
the initial atomic ratio of 1:2:2 is adopted from literature \cite{Ghadraoui}.
All the preparation procedures were carried out in an argon
protected glove box with the water and oxygen content below
0.1 ppm. The obtained samples are hard and quite stable in the
air.

Room temperature powder x-ray diffraction (XRD) measurements of
CeNi$_{2-\delta}$(As$_{1-x}$P$_{x}$)$_{2}$ were carried out on a
PANalytical x-ray diffractometer (Model EMPYREAN) with a
monochromatic Cu $K_{\alpha1}$ radiation. The chemical compositions
were verified by energy dispersion x-ray spectrometer (EDS)
affiliated to a field emission scanning electron microscope (FEI
Model SIRION). The electron beam was focused on a crystalline grain
and 3 EDS spectra from different grains were collected to avoid
randomicity. In-plane longitudinal electrical resistivity
$\rho_{xx}(T)$ was measured by the DC four-probe method in a $^3$He
cryostat down to 0.3 K (Oxford instrument, model Heliox VL). The
magnetic measurements were performed in a magnetic property
measurement system (Quantum Design, MPMS-5) with the temperature
range of $T$ = 2-400 K. The specific heat and transverse Hall
resistivity $\rho_{xy}(T)$ measurements were carried out in a
physical property measurement system (Quantum Design, PPMS-9) down
to about 0.5 K.

\section{Results and discussion}
\subsection{Sample characterizations}

\begin{figure}[t]
\centering
\includegraphics[width=7.5cm]{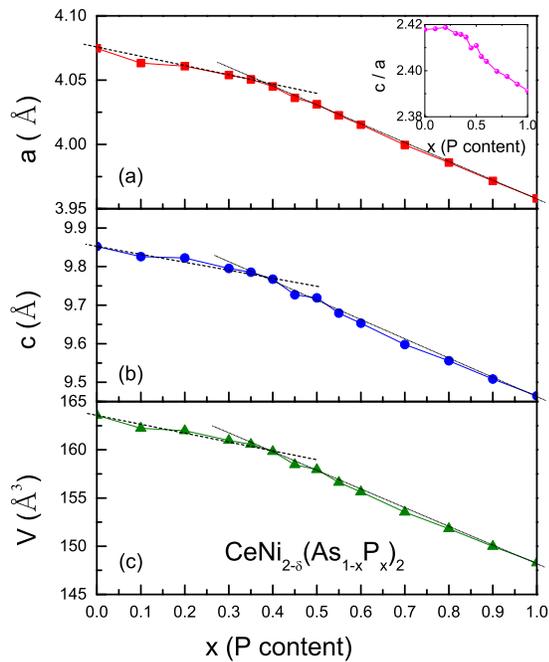}
\caption{(Color online) Lattice parameters $a$, $c$ and unit-cell
volume $V$ as a function of the P content $x$ in (a), (b) and (c).
The dashed and dotted lines are linear fit for $x$ = 0.0-0.4 and
0.5-1.0, respectively. Inset to (a) shows the ratio of $c/a$. \label{fig1}}
\end{figure}

The room temperature XRD measurements were performed on several
CeNi$_{2-\delta}$(As$_{1-x}$P$_{x}$)$_{2}$ samples. Almost all patterns can
be well indexed with the ThCr$_{2}$Si$_{2}$-type body centered
tetragonal structure with the space group $I4/mmm$ (No. 139), except
for a slight NiAs/Ni$_2$P non-magnetic impurity in samples with low/high
P-concentration (data not shown). The peaks shift to higher degree
with increasing P-concentration, consistent with the fact that the
ion radius of P is smaller than that of As. The lattice parameters
$a$, $c$ and the unit cell volume $V$, which are estimated by the
least square method using at least 25 reflections, are plotted in
Fig. \ref{fig1} as a function of the P content $x$. Two distinct, linear
regimes with different slopes are clearly discernible. All the
lattice parameters for $x$ $\geq$ 0.5 decrease faster than those
with lower concentration. This is an indication of intermediate
valence of cerium, since the ion radius of Ce$^{4+}$ is smaller than
that of Ce$^{3+}$. This is confirmed by the previous report
\cite{CeNi2P2} which shows that the Ce ion in CeNi$_{2-\delta}$P$_{2}$ is
in the intermediate valence state.

We also performed Rietveld refinement \cite{Rietveld} on the XRD
data of the parent compounds CeNi$_{2-\delta}$As$_{2}$ and
CeNi$_{2-\delta}$P$_{2}$ (data not shown). The lattice parameters
$a$ and $c$ are 4.0745(1) ${\AA}$ (3.95789(4) ${\AA}$) and 9.8514(3) ${\AA}$
(9.4650(1) ${\AA}$) for the As- (P-) compound, respectively, in agreement
with the previously reported values\cite{Luo2012,CeNi2P2}.  The derived
c/a ratio collapses from 2.418 to 2.391 (seen from inset to
Fig. \ref{fig1}(a)). This evidences an increased $c$-$f$ hybridization or
Kondo effect which will be discussed latter. The derived occupation
of Ni site is 0.78 and 0.93 for As- and P-end compound, respectively,
confirmed by the EDS analysis. Therefore, the number of Ni atom is
obtained as $\delta$ $\approx$ 0.07-0.22 for this series compounds.
From the EDS measurements, the molar ratio of Ce:Ni in all the series
compounds is about 1:(1.71$\pm$0.15), while the ratio of P to As
is almost equal to the nominal value. So the Ni vacancies exist in
both end compounds. Moreover, the sum of P and As contents is
slightly higher than the nominal value. We note that a similar
observation was reported in a number of other arsenic compounds\cite{Bao}.

\subsection{Magnetic susceptibility and Hall resistivity}

\begin{figure*}[t]
\centering
\includegraphics[width=12cm]{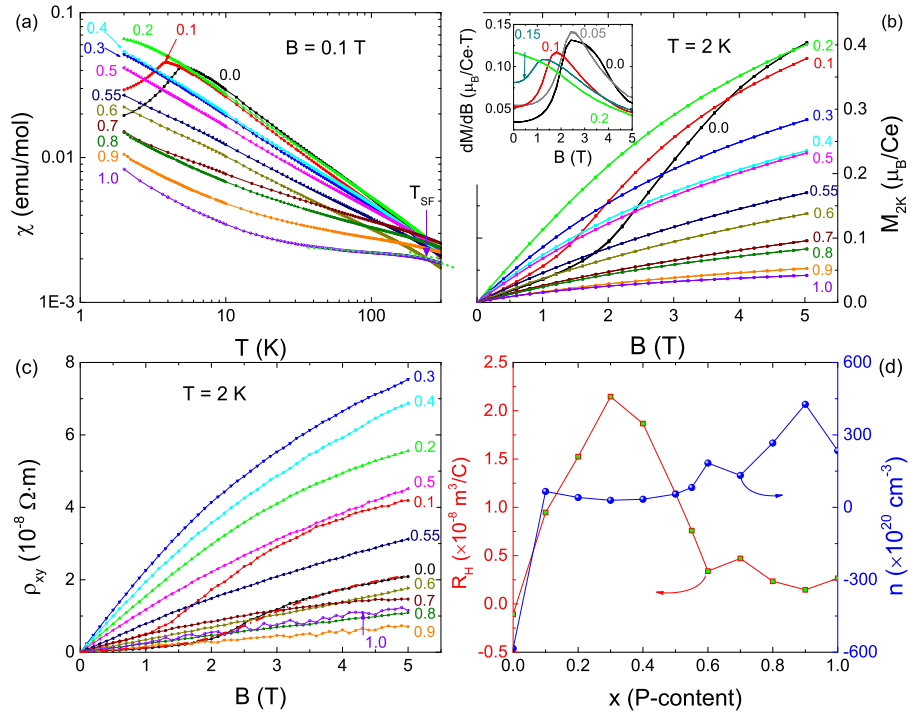}
\caption{(Color online) $\chi_{mag}(T)$ (a) and $M_{2K}(B)$ (b) of
CeNi$_{2-\delta}$(As$_{1-x}$P$_{x}$)$_{2}$ for the selected concentrations.
(c) Magnetic field dependence of the Hall resistivity, $\rho_{xy}(B)$,
measured at $T$ = 2 K. The dashed line presents the fitted curve for
$x$ = 0.0 using Eq.\ref{eq1} (see the text).
(d) The $x$ evolution of the Hall coefficient $R_{H}$ and the carrier
density $n$ on the left and right panel, respectively.
Inset to (b) presents d$M$/d$B$ versus $B$ at $T$ = 2 K for $x$ = 0.0-0.2.
\label{fig2}}
\end{figure*}

Fig. \ref{fig2}(a) presents the temperature dependence of the molar magnetic
susceptibility $\chi_{mag}$ for eleven CeNi$_{2-\delta}$(As$_{1-x}$P$_{x}$)$_{2}$
samples of different $x$ measured at $B$ = 0.1 T on a logarithmic plot.
Here the magnetic contribution is derived from $\chi_{mag}(T)$ =
$\chi_{Ce}$ - $\chi_{La}$, where $\chi_{La}$ is the magnetic susceptibility of
isostructural nonmagnetic compound LaNi$_{2}$As$_{2}$ (data not shown).
Several features are clearly revealed: (i) A kink at $T_{N}^{\chi}$ = 4.8 K,
the characteristic of a second-order AFM transition, is pronounced
for CeNi$_{2-\delta}$As$_{2}$. As the amount of P is increased, the AFM
transition shifts to lower temperature and falls below our base
temperature for $x$ $\geq$ 0.2. (ii) The temperature dependence of
magnetic susceptibility, which is strong for the As-end compound,
becomes rather weak with increasing $x$. Besides, the isothermal
magnetization measured at $T$ = 2 K decreases almost monotonously
with increasing $x$ as shown in Fig. \ref{fig2}(b), also suggesting the
screening of Ce local moments, or delocalization of the $f$ electrons.
To interpret this tendency qualitatively, the $\chi(T)$ curves
at high temperature are fitted with a modified Curie-Weiss
law, $\chi$ = $\chi_{0}$ + ${\cal C}/(T-\theta)$. Here $\chi_{0}$ is
a temperature independent susceptibility from the core diamagnetism,
the van Vleck and Pauli paramagnetism, ${\cal C}$ is the Curie
constant and $\theta$ is the Curie temperature. The effective moment
$\mu_{eff}$, which could be derived from the Curie constant, is then
estimated to be around 2.26$\mu_{B}/$Ce, showing a very weak
variation with $x$ for $x$ $<$ 0.5. Here we recall that usually the
3$d$ electrons of Ni ions do not show magnetism in the nickel based
pnictides\cite{Luo2012,nickel},  so the magnetic moments in
CeNi$_{2-\delta}$(As$_{1-x}$P$_{x}$)$_{2}$ should come from the Ce
4$f^1$-electrons. The observed value of the effect moment is close to
but slightly smaller than that of the free Ce$^{3+}$ ion, 2.54$\mu_{B}$.
The slight reduction of the Ce moments is usually ascribed to the
crystalline electric field (CEF) effect\cite{Luo2012}.  With increasing
P content, $\mu_{eff}$ decreases to 1.88$\mu_{B}/$Ce for $x$ = 0.9,
confirming the enhanced itineracy of 4$f$-electrons due to the
strengthened coherent Kondo screening. (iii) As P concentration is
increased to $x$ = 0.7, a broad hump in susceptibility appears
around 100 K, corresponding to the so-called spin fluctuation
temperature $T_{SF}$\cite{1}.  The position of this hump shifts to
higher temperature and becomes more apparent as $x$ increases. In
CeNi$_{2-\delta}$P$_{2}$, the Curie-Weiss law mentioned above is
abided for $T$ $<$ 150 K, with a much reduced value of $\mu_{eff}$
= 0.34 $\mu_{B}/$Ce, while this law is violated in a high temperature
range of $T$ = 200-400 K. These results reveal the intermediate
valence behavior for the compounds with $x$ $\gtrsim$ 0.7. Moreover,
the characteristic temperatures $T_{SF}$ and $T_{coh}$, where
$\chi_{mag}(T)$ and $\rho_{mag}(T)$ exhibit respective local maxima,
are located at almost the same temperature as shown by arrows in Fig.
\ref{fig2}(a) and Fig. \ref{fig4}(d). This observation favors the
interpretation of $T_{coh}$ in terms of spin-scattering mechanism
in the intermediate valence state\cite{1,Varma76}.

The isothermal magnetizations $M(B)$ for
CeNi$_{2-\delta}$(As$_{1-x}$P$_{x}$)$_{2}$, measured at $T$ = 2 K
in a $B$-sweep mode, is shown in Fig. \ref{fig2}(b). For the parent
compound CeNi$_{2-\delta}$As$_{2}$, $M(B)$ displays a step-like
behavior around 2.3 T, which is ascribed to a field-induced MMT\cite{Luo2012}.
 In order to determine the MMT field $B_{m}$ exactly,
the derivative of magnetic field for the magnetization, d$M$/d$B$,
is calculated as shown in the inset to Fig. \ref{fig2}(b). $B_{m}$
is thus obtained at the peak of d$M(B)$/d$B$. As $x$ is increased,
$B_{m}$ shifts from 2.3 T for $x$ = 0.0 to a lower field of 1.21 T
for $x$ = 0.15 and is indiscernible for $x$ $\geq$ 0.2 in the
temperature limit of our measurement. It$'$s noted that the
corresponding hysteresis around $B_{m}$ of the first-order nature
is indiscernible. This is probably because the magnetic correlations
among cerium moments are strongly anisotropic; thus, the hysteresis
detected only for $B||c$ of single crystalline samples in Ref.\cite{Luo2012}
is smeared out in the polycrystalline samples studied here.

We also measured the transverse Hall resistivity at $T$ = 2 K as a
function of the applied magnetic field $\rho_{xy}(B)$ as shown in
Fig. \ref{fig2}(c). Interestingly, the field dependence of $\rho_{xy}$
resembles the magnetization, $M$, with the following features: (i)
the nonlinearity in the full magnetic field range of $B$ = 0-5 T for
the whole system; (ii) a pronounced step-like increase at $B_{m}$ =
2.3 T for dopants with $x$ $\leq$ 0.1 without obvious hysteresis.
The critical field $B_{m}$ for the spin-flop MMT shifts to lower
fields as the P-concentration is increased. It is noteworthy that
the evolution of the magnetic field vs. P-doping concentration is
remarkably similar to that of $B-p$ relationship drawn in Ref.\cite{Luo2015}.

With these observations,
we fit
 the measured results
 for $\rho_{xy}(B)$
 by the following expression:
\begin{equation}
\rho_{xy}(B)= R_{H}B+R_{S}\mu_{0}M(B).\label{eq1}
\end{equation}
The first term represents the normal Hall effect originated from the
Lorentz force, while the second term represents the anomalous Hall
effect (AHE)  originated presumably from the magnetic fluctuations
of the localized Ce 4$f$-spins, the skew scattering\cite{Hall}.  The
derived Hall coefficient $R_{H}$ and the carrier density $n$, using
a simple single band model according to the relationship of $R_{H}$
$\propto$ $1/n$, are presented in Fig. \ref{fig2}(d) for various $x$.
The sign of $R_{H}$ ($n$) changes from negative for $x$ = 0.0 to
positive for the rest of this family, while the anomalous Hall contribution
$R_{s}$ is always positive. In particular, the magnitude of $n$ is quite small,
showing a significant reduction in the range of
0.1 $\leq x \leq$ 0.7, with a slight increase in the compounds on
the P-rich side.
The lowest magnitude is about one order smaller than
about 30$\times$10$^{20}$ cm$^{-3}$ for the end compound $x$ = 1.0.
This is consistent with the presence of the Ni vacancies mentioned
previously. As we will show later, the low carrier density due to
the Ni vacancies may account for the semimetal behavior in the
As-rich compounds, and the slight increase of $n$ in the P-rich side
compounds may also explain the crossover from semimetal-like to
metallic behavior at low temperature. It is, therefore, speculated
that P-doping has both shifted the Fermi level and changed the
dominant carrier from electron to hole. The Hall resistivity data
for CeNi$_{2-\delta}$As$_{2}$ at different temperatures are also
fitted using the above equation \ref{eq1}. $R_{H}$ increases to a broad
maximum around $T_{coh}$ before decreasing sharply to a small value
in the low temperature coherent-band regime (data not shown). This
behavior is reminiscent of CeAl$_{3}$, CeRu$_{2}$Si$_{2}$, and
many other heavy fermion compounds\cite{AHE}.

\subsection{Specific heat}

\begin{figure}[t]
\centering
\includegraphics[width=8.3cm]{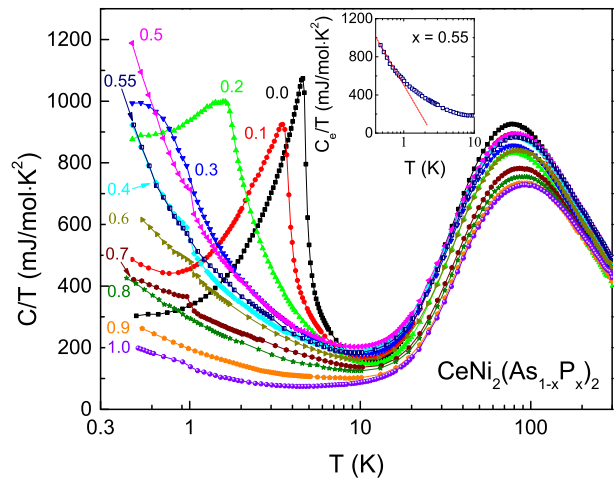}
\caption{(Color online) Temperature dependence of specific heat divided
by temperature, $C/T$, for CeNi$_{2-\delta}$(As$_{1-x}$P$_{x}$)$_{2}$.
Inset displays the electronic contribution $C_{e}(T)/T$ for $x$ = 0.55
at $T$ = 0.4-10 K. The dashed line is a linear fit. \label{fig3}}
\end{figure}

The specific heat coefficient of
CeNi$_{2-\delta}$(As$_{1-x}$P$_{x}$)$_{2}$, $C(T)/T$, is plotted
in the main panel of Fig. \ref{fig3} in semi-logarithm scale.
A prominent feature for the As-rich compounds is the $\lambda$-type
kink, typical of a second-order AFM transition. This peak is most
pronounced in the parent compound, exhibited at 4.8 K in consistent
with the previously reported value\cite{Luo2012}.  The magnetic
transition temperature, $T_{N}^{C}$, can then be determined from the
derivative of $C(T)/T$ at the sharp kink. For $x$ = 0.0-0.4, the
peak broadens and moves to lower temperature with increasing the
P-content $x$. For $x$ = 0.5, $C(T)/T$ increases drastically down
to our base temperature while the derivative curve still drops down
sharply. This is a characteristic feature of quantum criticality,
implying a zero temperature QCP near $x=0.5$. Moreover, the
electronic contribution $C_{e}(T)/T$ at $x$ = 0.55, derived by
subtracting the phonon term $\beta T^2$, increases in a logarithmic
scale below $\sim$ 1 K, shown in the inset to Fig. \ref{fig3}. This
typical NFL behavior provides strong evidence for a divergent
quasiparticle mass near the QCP. Far away from the critical point,
the divergent tendency is suppressed to some extent, and finally,
$C(T)/T$ tends to saturate at low temperature for $x$ = 1.0,
indicating the recovery of the paramagnetic Fermi liquid state.
%The moderately enhanced values of effective mass for $x$ = 0.6-1.0
%confirm the intermediate valence state with strong $c$-$f$ coupling.
Note that there is another anomalous peak in $C(T)/T$ around 1 K for
several samples. This anomaly probably comes from tiny amounts of
unknown impurity phase, since its position does not change with P
doping and nothing is observed from electrical resistivity curves
over this temperature range. The upturn below 0.7 K in the $x$ = 0.1
sample may be due to a nuclear quadrupolar Schottky anomaly arising
mainly from the As atom.

In order to characterize the variation of the Kondo coupling, or
more precisely the hybridization between Ce-4$f$ and the conduction
electrons upon P doping, we also consider the variation of the
Sommerfeld coefficient $\gamma_{0}$ with increasing $x$. Except for
the parent compound with $x$ = 0.0, $\gamma_{0}$ is extracted from
$C/T$ data down to the base temperature ($T$ = 0.45 K) since a
divergent scattering is driven mainly by strong spin fluctuations
around the QCP; besides, Kondo effect is expected to dominate over
the RKKY interaction for $x$ $\geq$ 0.6 samples. In Fig. \ref{fig5},
both Sommerfeld coefficient $\gamma_{0}$ and the residual resistivity
$\rho_{0}$ show significant peaks at $x$ = 0.5, in further support
of a QCP near this doping level. We could expect that the peaks in
$\gamma_{0}$ and $\rho_{0}$ will be more divergent as temperature
is further reduced forwards zero.

\subsection{Longitudinal electrical resistivity}

\begin{figure*}[t]
\centering
\includegraphics[width=12.5cm]{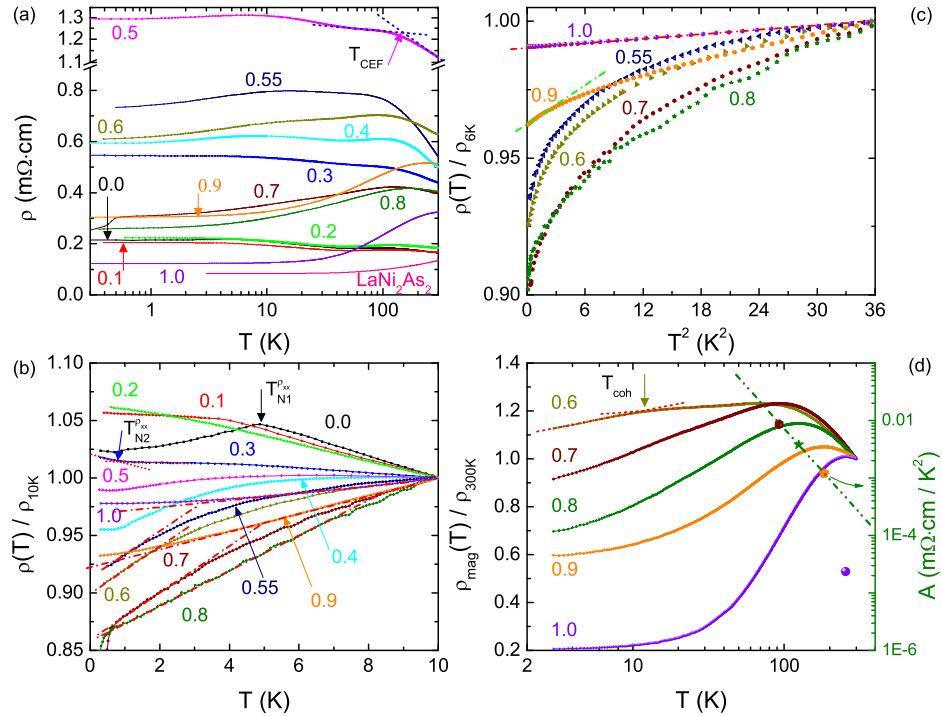}
\caption{(Color online) (a) Temperature dependence of the electrical
resistivity $\rho_{xx}(T)$ for CeNi$_{2-\delta}$(As$_{1-x}$P$_{x}$)$_{2}$
series as well as LaNi$_{2}$As$_{2}$ in a log-log scale.
The normalized resistivity for the selected concentrations is shown as
$\rho_{xx}$/$\rho_{10K}$ versus $T$ below 10 K in (b) and $\rho_{xx}$/$\rho_{6K}$
versus $T^2$ below 6 K in (c), respectively.
The dashed-dotted lines are linear fit to $\rho_{xx}$.
$T_{N2}^{\rho_{xx}}$ is determined by the intercrossing of two dotted lines.
The normalized resistivity $\rho_{xx}(T)$/$\rho_{300K}$ at $T$ = 2-400 K
on the left axis of (d) for $x$ = 0.6-1.0 samples.
The log-log plot of $A$ coefficient versus $T_{coh}$ on the right axis of (d).
The dashed-dotted-dotted line indicates $A$ $\propto$ $(T_{coh})^2$.
$T_{CEF}$ ($T_{coh}$) is shown by an arrow for $x$ = 0.5 sample in (a) ($x$ = 0.6 in (d)).
\label{fig4}}
\end{figure*}

The electrical resistivity $\rho_{xx}(T)$ in the full-$T$ range and
the normalized resistivity $\rho_{xx}(T)$/$\rho_{10K}$ below 10 K
of CeNi$_{2-\delta}$(As$_{1-x}$P$_{x}$)$_{2}$ are shown in
Fig. \ref{fig4}(a) and (b), respectively. For the parent compound
CeNi$_{2-\delta}$As$_{2}$, the AFM ordering phase is entered below
$T_{N1}^{\rho_{xx}}$ = 4.88 K, directly visible by the pronounced
cusp in $\rho_{xx}(T)$. A negative logarithmic slop can be seen
above $T_{N1}^{\rho_{xx}}$ up to 40 K, indicating the Kondo effect.
A broad hump develops at higher temperature $T_{CEF}$ = 105 K which
is due to the Kondo scattering on the excited CEF levels. When the P
content increases,  $T_{N1}^{\rho_{xx}}$ shifts to lower temperature,
as reflected by an increasing rounding in $\rho_{xx}(T)$. For $x$
$\geq$ 0.2, the increase in $\rho_{xx}(T)$, namely an ¡°inflection¡±,
$T_{N2}^{\rho_{xx}}$, determined by the intercrossing of two dotted
lines (arrow in Fig. \ref{fig4}(b)), signals the onset of AFM
ordering. This inflection has been observed in CeCu$_{6-x}$Au$_{x}$
and many other Kondo lattice systems below $T_{N}$\cite{Lohneysen98}.
Note that $T_{N2}^{\rho_{xx}}$ is further suppressed to 0.45 K for
$x$ = 0.5 and invisible for $x$ $\geq$ 0.55 in our measurement
limit.

\begin{figure}[t]
\centering
\includegraphics[width=8.8cm]{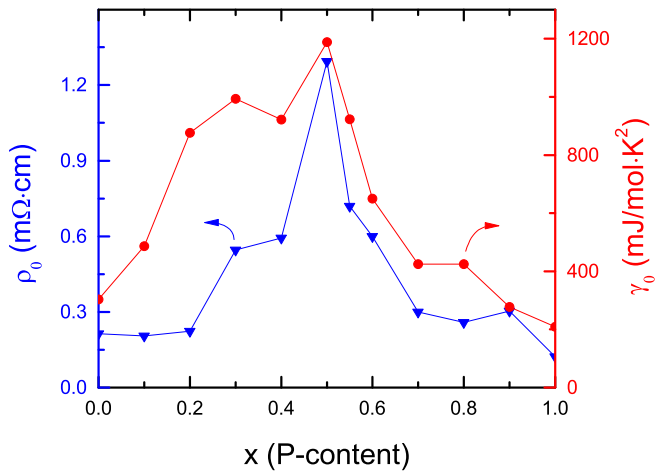}
\caption{(Color online) The evolution of residual resistivity $\rho_{0}$ and electronic Sommerfeld
coefficient $\gamma_{0}$ versus $x$ on the left and right axis, respectively.
\label{fig5}}
\end{figure}

A remarkable feature for 0.0$\leq$ $x\leq$ 0.5 is that $\rho_{xx}(T)$
increases slowly with decreasing temperature, showing the semimetal
behavior at very low temperatures.  While for $x$ $\geq$ 0.6, the
metallic behavior is exhibited below $T_{coh}$, determined as the
local maxima shown in Fig. \ref{fig4}(d). For $T<T_{coh}$,
$\rho_{xx}(T)$ decreases monotonically with decreasing temperature,
indicating the regime for the coherent Kondo screening. Note that in
the P-rich case, the enhancement of electrical conductivity is
clearly reflected by the decreased residual resistivity and increased
carrier density $n$, shown in Fig. \ref{fig5} and Fig. \ref{fig2}(d),
respectively, for higher doping concentrations. All these manifest
the enhanced itineracy of Ce 4$f$-electrons with increasing $x$ in
the P-rich case. Moreover, a nearly linear temperature dependence in
$\rho_{xx}(T)$ with varying slopes could be observed in the
low-temperature regime for $x$ = 0.55-0.8. Such linear resistivity
behavior could be also extracted for $x$ $\leq$ 0.9 over a limited
range of $T$ as shown by the dashed-dotted lines in Fig. \ref{fig4}(b).
The lower end of that temperature range, denoted by $T_{FL}$, signals
the onset of the Fermi liquid state. The estimated $T_{FL}$
increases from 0.95 K for $x$ = 0.9 to 5.5 K for $x$ = 1.0.
Therefore, we further fit the low temperature resistivity in terms
of a power law $\rho_{xx}(T)$ = $\rho_{0}$ + $AT^\alpha$, for $T$
$\leq$ 3 K. Such fitting is carried out only for $x$ $\geq$ 0.55
because of the semimetal behavior when $x$ $\leq$ 0.5. The coefficient
$A$ is related to the effective mass of charge carriers through
$A$ $\sim$ ($m^{*}$$/$$m_{0})^2$. As P content is increased, the
coefficient $A$ is strongly decreased, and $\alpha$ varies from
0.67 for $x$ = 0.55 to 2.41 for $x$ = 1.0. The residual resistivity
$\rho_{0}$, determined at $T$ = 0.5 K, is plotted on the left panel
of Fig. \ref{fig5} as a function of $x$. The maximum value of
$\sim$ 1.3 m$\Omega\cdot$cm for $x$ = 0.5 indicates strong quantum
fluctuations near the critical point, although high density of voids
and/or microcracks can not be fully neglected. The residual resistivity
$\rho_{0}$ for the As- and P-end compounds is likely due to
the Ni-deficiency and$/$or the relatively low carrier density $n$.
Note that the magnitude of $\rho_{0}$ of these Ni-based pnictides,
falling in the order of m$\Omega\cdot$cm, is comparable to some
of the iron-based pnictides\cite{Pnictide}.

In order to extract the magnetic contribution from the Ce
4$f$-electrons, $\rho_{mag}$ is introduced by subtracting the
corresponding resistivity of LaNi$_{2}$As$_{2}$ which is
presented in Fig. \ref{fig4}(a). The temperature dependence of
$\rho_{mag}(T)$, normalized by the value at $T$ = 300 K, is plotted
in the left axis in Fig. \ref{fig4}(d) (for $x$ $\geq$ 0.6).
The $\rho_{xx}(T)$ curves show
different local maxima, $T_{CEF}$ and $T_{coh}$; their values are
determined from the intercrossing of two dashed lines as indicated
in Fig. \ref{fig4}(a) and Fig. \ref{fig4}(d). Roughly speaking,
$T_{CEF}$ corresponds to the Kondo scattering on the ground state
and the excited CEF levels, while $T_{coh}$ reflects the onset of
a coherent state. $T_{CEF}$ depends weakly on $x$, and decreases
slightly in the P-rich side only. Because of the semimetal behavior
for $x$ $\leq$ 0.5, $T_{coh}$ develops when $x$ $\geq$ 0.6 in the
P-rich regime where the magnetic moments are fully screened. In our
measurements, $T_{CEF}$ and $T_{coh}$ merge at $x$ $\approx$ 0.65,
after that an intermediate valence regime is entered where the
Kondo coherence is expected to become larger than the CEF splitting
energy. We plot the coefficient $A$ vs. $T_{coh}$ in a log-log scale
on the right panel of Fig. \ref{fig4}(d). The curve follows the
expected relationship of $A$ $\propto$ $(T_{coh})^{-2}$  for 0.7
$\leq$ $x$ $\leq$ 0.9 (see the dashed-dotted-dotted line in Fig.
\ref{fig4}(d)), but deviates significantly as $x$ is larger than 0.9,
suggesting a valence crossover induced by the P-doping. More
evidence, such as x-ray absorption, will be instructive to further
characterize the valence state.

For $x$ = 0.7, a sudden drop of about 15$\%$ in $\rho_{xx}(T)$
is presented below $T$$'$ = 0.51 K, seen in Fig. \ref{fig4}(b). We
find that this silent anomaly can be suppressed by applying either
a magnetic field of $B$ = 0.15 T or a large amount of current of
$I$ = 4 mA. Traces of such effect are also observed for $x$ = 0.6
and 0.8, of which the resistivity starts to decrease though not as
strong as that of $x$ = 0.7.
%Note that no magnetic ordering is found in samples with
%$x$ $\geq$ 0.55 from the specific heat measurement.
It might be a signal for the system entering a coherent
superconducting phase in the very low temperature region for these
samples ($x$ = 0.6-0.8). However, this possibility can not be
concluded in our present measurements since a superconducting
impurity phase can not be fully excluded.
Further measurements down
to much lower temperatures and high quality single crystals are
highly desirable to clarify this anomaly.

\subsection{Discussion}

\begin{figure}[t]
\centering
\includegraphics[width=8.5cm]{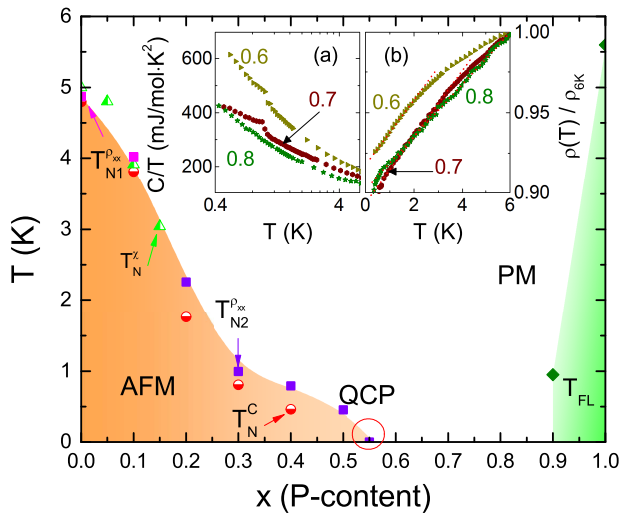}
\caption{(Color online) Phase diagram of CeNi$_{2-\delta}$(As$_{1-x}$P$_{x}$)$_{2}$
as a function of P concentration $x$. Solid squares, half-filled triangles,
and half-filled circles correspond to the AFM transition temperatures from
different property measurements: electrical resistivity, magnetic susceptibility
and specific heat. Fermi liquid state is recovered below $T_{FL}$ which is shown
as filled diamonds. Around $x_{c}$ = 0.55, QCP is marked by a circle.
Inset (a) and (b) respectively show the specific heat divided by temperature,
$C/T$ versus $\log$($T$), and the normalized electrical resistivity,
$\rho_{xx}(T)$/$\rho_{6K}$, below 6 K for $x$ = 0.6, 0.7 and 0.8.
The dashed lines are linear fits to $\rho_{xx}(T)$.
\emph{The dotted line is a guide to the eyes.}
\label{fig6}}
\end{figure}

The deduced phase diagram for CeNi$_{2-\delta}$(As$_{1-x}$P$_{x}$)$_{2}$
is presented in the main panel of Fig. \ref{fig6} in terms of P
doping concentration $x$. The value of $T_{N1}^{\rho_{xx}}$
($T_{N2}^{\rho_{xx}}$) is slightly higher than $T_{N}^C$ from the bulk
measurements, especially for samples with 0.2 $\leq$ $x$ $\leq$ 0.5.
Nevertheless, the N\'{e}el temperatures determined from different
techniques are in reasonable agreement with each other. The isovalent
substitution of P for As in CeNi$_{2-\delta}$(As$_{1-x}$P$_{x}$)$_{2}$
results in lattice contraction which in turn leads to the enhancement
of Kondo coupling $J$ between the conduction electrons and the 4$f$
moments. Consequently, the long range AFM order is suppressed
continuously by the Kondo coupling or $c$-$f$ hybridization for
$x$ $\leq$ $x_{c}$. Within the resolution of our measurements, the
critical P concentration $x_c$ is close to 0.55 (more precisely,
somewhere between 0.5-0.6). Paramagnetic state is entered as $x$
is larger than 0.55. The itineracy of Ce 4$f$-electron is enhanced
as evidenced from the reduction of the effective moment. Very close
to the magnetic instability $x_{c}$ = 0.55, the specific heat
coefficient $C/T$ diverges logarithmically, $C/T$ $\propto$
-$\log(T)$, and the electrical resistivity varies linearly,
$\rho_{xx}$ $\propto$ $T$. Both of them are typical NFL behavior
that appears in the vicinity of a QCP. Note that around the QCP,
there is no hysteresis loop in $\rho_{xx}(T)$ and $C(T)/T$,
implying the second-order nature of the quantum phase transition.
Since the valence instability is well separated from the fluctuations
of the magnetic order parameter, the quantum criticality is of
magnetic nature. Slightly above the QCP, at $x$ $\approx$ 0.65, the
Kondo temperature and the CEF energy become compatible: $T_{coh}$
$\approx$ $T_{CEF}$. This is reminiscent of the overlap regime in
the phase diagrams of Ce122 heavy fermion superconductor family\cite{CeCu2Si2,CeCu2Ge2,CeAu2Si2}.

The P-end parent compound CeNi$_{2-\delta}$P$_{2}$ shows the typical
Fermi liquid behavior: $\rho_{xx}$ $\sim$ $A T^{2.2}$, and $C/T$
saturates in the low temperature limit with moderately enhanced
$\gamma_0$ $\approx$ 209 mJ/mol$\cdot$K$^2$. The Kadowaki-Woods
ratio $A/\gamma^2$ is then estimated to be 4.41$\times$10$^{-7}$
$\mu\Omega\cdot$cm(K$\cdot$mol/mJ)$^2$. This value is more
comparable to the generalized Kadowaki-Woods relation
$A/\gamma^2/[N(N-1)/2]$ = 6.7$\times$10$^{-7}$
$\mu\Omega\cdot$cm(K$\cdot$mol/mJ)$^2$ for the Ce-based Kondo
lattice compounds with a large orbital degeneracy $N$, rather than
the ratio for the standard Fermi liquid with a Kramers doublet
ground state\cite{K-W1, K-W2}.  These properties suggest that
CeNi$_{2-\delta}$P$_{2}$ has an almost fully degenerate ground
state due to the largely enhanced $c$-$f$ hybridization compared
with the CEF splitting energy. This is further supported by the
decreasing tendency: $A/\gamma^2$ = 4.59$\times$10$^{-5}$ for $x$
= 0.6 and $A/\gamma^2$ = 1.51$\times$10$^{-5}$
$\mu\Omega\cdot$cm(K$\cdot$mol/mJ)$^2$ for $x$ = 0.9.

It is interesting to compare the chemical pressure effect in the
series CeNi$_{2-\delta}$(As$_{1-x}$P$_{x}$)$_{2}$ with the hydrostatic
pressure effect in CeNi$_{2-\delta}$As$_{2}$ based on the respective
phase diagrams. Doping with 55$\%$ P leads to a contraction of the
unit-cell volume, $\Delta$$V/V_{0}$ = (V$_{0}$-V$^{\prime}$)/V$_{0}$,
of 4.3$\%$. This volume contraction corresponds to that of
CeNi$_{2-\delta}$As$_{2}$ under the critical pressure of $P_{c}$ =
2.7 GPa\cite{Luo2015}.  A linear extrapolation towards the 100$\%$
P-doping case would suggest that a volume contraction of $\approx$
9.4$\%$ corresponds to an estimated pressure of $P$ $\approx$
6.0 GPa in order to recover the Fermi liquid state. This is
over-estimated, since the lattice cell volume does not follow a
Vegard$'$s law between As- and P- rich compounds, but decreases
more rapidly beyond 40$\%$. While there is no simple one-to-one
quantitative correspondence between the chemical doping
and physical pressure experiments, the phase diagrams deduced
respectively from chemical doping and hydrostatic pressure show
considerable similarities. Both indicate the existence of an
AFM QCP that is accompanied by
a delayed coherent Kondo screening associated with the low carrier density.
The evidence for
quantum criticality  \cite{Hertz, Moriya, Millis,
local,Gegenwart}
is clear both
at $x_{c}$ in the present study and $p_{c}$
in Ref.\cite{Luo2015}.
Moreover, the similarity in the quantum critical properties, such as the $T$-logarithmic divergence
in $C_{e}/T$, at the chemical-pressure-induced $x_{c}$ here and at the pressure-driven
$p_{c}$ in Ref.\cite{Luo2015} makes a strong case for the intrinsic nature of the quantum
criticality.

At the same time,
we also identify a key difference. In contrast to
the phase diagram of the parent compound CeNi$_{2-\delta}$As$_{2}$
as a function of hydrostatic pressure\cite{Luo2015},  here we
find a surprising result in the $T-x$ phase diagram of
CeNi$_{2-\delta}$(As$_{1-x}$P$_{x}$)$_{2}$ that the NFL
behavior persists over a nonzero range of doping concentration,
$x_c$ $<x<$ 0.9. As seen as inset (a) to Fig. \ref{fig6}, the
specific-heat coefficient of $x$ = 0.6-0.8 samples is
logarithmically
divergent at the lowest measured temperature range, $C/T$
$\varpropto$ -$\log$($T$). The temperature dependence of the
electrical resistivity, as shown in the inset (b) to Fig. \ref{fig6},
behaves linearly. The unique NFL behavior is also manifested in
the evolution of the power exponent $\alpha$, deduced from the
power law fitting described in section D. This exponent $\alpha$
increases slightly with increasing $x$ from $\alpha$ = 0.67 for
$x$ = 0.55 to 1.18 for $x$ = 0.8 but rapidly increases to be
larger than 2 for $x$ $\sim$ 1.0, signaling the Fermi liquid
behavior. We note that in our measurements the Hall coefficient,
$R_{H}$, varies smoothly with $x$ across the QCP, as shown as
squares in Fig. \ref{fig2}(d). This behavior is in contrast to
the related oxypnictides CeNiAs$_{1-x}$P$_x$O in which $R_{H}$
reveals a drastic change across the QCP\cite{CeNiAsO}.  The
evolution of $R_{H}$ in our system may be delayed by the broadened
NFL region since
$n$ increases
significantly for $x$ = 0.7-0.9.
However, we still cannot conclude the absence of a drastic
change from a smaller 4$f$-localized Fermi surface to a larger
4$f$-itinerant one, since these data are measured at $T$ = 2 K.
These properties suggest the presence of NFL behavior over
a finite zero-temperature region of $x$ = 0.6-0.8 in
CeNi$_{2-\delta}$(As$_{1-x}$P$_{x}$)$_{2}$. This behavior has been
reported in the field-tuned transition for Ir-doped \cite{Ir-Yb122}
and Ge-doped \cite{Ge-Yb122} YbRh$_{2}$Si$_{2}$.
Our results indicate a
NFL phase which resembles the trajectory "III" in the global
phase diagram of the heavy fermion metals\cite{global,SiPaschen}.
Different from the universal SDW-type scenario, i.e., trajectory
"II" \cite{Hertz, Moriya, Millis} and the unconventional quantum
criticality, i.e., trajectory "I", which incorporates not only
the slow fluctuations of the AFM order parameter but also the
energy scale $E^{*}$ associated with the breakup of the Kondo
singlet\cite{local,Gegenwart} , this route describes the
quantum phase transition
from a Kondo-destroyed antiferromagnetic phase with a small Fermi surface
(AF$_{\rm{S}}$) to a paramagnetic heavy fermion phase with a large Fermi surface
(P$_{\rm{L}}$)
 through the intermediate
Kondo-destroyed paramagnetic phase with a small Fermi surface
(P$_{\rm{S}}$)
phase.
Our results set the stage for measurements down to
lower temperatures
and theoretical analysis on the band structure and Fermi surface
topology
for further elucidation of
the nature of the QCP in this
intriguing
class of
pnictide-based heavy fermion
materials.

\section{CONCLUSION}

In summary, the magnetization, specific heat and transport
measurements have been performed on
CeNi$_{2-\delta}$(As$_{1-x}$P$_{x}$)$_{2}$ ($\delta$ $\approx$
0.07-0.22) system, and a global phase diagram is drawn as a
function of P-doping concentration $x$. Evidence of low carrier
density is found for $0.1 \leq x\leq 0.7$ from Hall resistivity
measurement. The N\'{e}el temperature $T_{N}$ is suppressed
continuously upon increasing $x$ which is invisible for $x_{c}$
= 0.55, the QCP. In the vicinity of QCP, the NFL behavior is
manifested by $C/T$ $\propto$ -$\log(T)$ and $\rho_{xx}$ $\propto$
$T$, and a divergent effective carrier mass is evidenced
from $\rho_{0}$ and $\gamma_{0}$. We find the surprising result
that the NFL behavior of the specific heat and electrical
resistivity persists over a nonzero range of doping concentration
for $x_c<x<0.9$; the Fermi liquid behavior is not recovered
until $x \geq 0.9$. We discuss the properties of
CeNi$_{2-\delta}$(As$_{1-x}$P$_{x}$)$_{2}$ together with those
of its 1111 counterpart, CeNi(As$_{1-x}$P$_{x}$)O, and draw
implications for the global phase diagram of the heavy fermion
metals. Our present study thus offers a new candidate material
for studying the universality classes of quantum criticality,
and highlights the effect of the low density of conduction
electrons in the nickel-based pnictides.

\begin{acknowledgments}

The authors acknowledge useful discussions with Y. K. Luo, S.
Paschen and J. D. Thompson. This work was supported in part by the
Ministry of Science and Technology of China  (Grant Nos.
2016YFA0300402 and 2014CB921203) and the National Natural Science
Foundation of China (Grant Nos. U1332209 and 11474082). Work at
Rice University was supported by the ARO Grant No.\
W911NF-14-1-0525 and the Robert A.\ Welch Foundation Grant No.\
C-1411, with travel support provided by the NSF Grant No.\
DMR-1611392.
\end{acknowledgments}


\begin{thebibliography}{10}
%%\bibitem{} %% \bibitem{label}

\bibitem{1} J. M. Lawrence, P. S. Riseborough, and R. D. Parks, Rep. Prog. Phys. \textbf{44}, 1 (1981).
\bibitem{2} G. R. Stewart, Rev. Mod. Phys. \textbf{73}, 797 (2001); \textbf{78}, 743 (2006).
\bibitem{3} H. von L\"{o}hneysen, A. Rosch, M. Vojta, and P. W\"{o}lfle, Rev. Mod. Phys. \textbf{79}, 1015 (2007).
\bibitem{Doniach} S. Doniach, Physica B+C \textbf{91}, 231-234 (1977).
\bibitem{4} F. Steglich, J. Arndt, O. Stockert, S. Friedemann, M. Brando, C. Klingner, C. Krellner, C. Geibel, S. Wirth, S. Kirchner, and Q. Si, J. Phys.: Cond. Matt. \textbf{24}, 294201 (2012).
\bibitem{5} Q. Si, F. Steglich. Science, \textbf{329}(5996), 1161-1166 (2010).
\bibitem{Nozieres98}P. Nozi\`{e}res, Eur. Phys. J. B \textbf{6}, 447 (1998).
\bibitem{Luo2012} Y. Luo, J. Bao, C. Shen, J. Han, X. Yang, C. Lv, Y. Li, W. Jiao, B. Si, C. Feng, J. Dai, G. Cao, and Z. A. Xu, Phys. Rev. B \textbf{86}, 245130 (2012).
\bibitem{Luo2015} Y. K. Luo, F. Ronning, N. Wakeham, X. Lu, T. Park, Z. A. Xu, and J. D. Thompson, Proc. Natl. Acad. Sci. USA \textbf{112}, 13520-13524 (2015).
\bibitem{Feng2016}
X.-Y. Feng, H. Zhong, J. Dai and Q, Si, arXiv:1605.02380.
\bibitem{CeNi2P2} H. Suzuki, H. Abe, H. Kitazawa, and D. Schmitt, J. Alloys Comp. \textbf{323-324}, 520-523 (2001).
\bibitem{Ghadraoui} E. H. El Ghadraoui, J. Y. Pivan, R. Gu\'{e}rin, O. Pena, J. Padiou, and M. Sergent, Mat. Res. Bull. \textbf{23}pp, 1345-1354 (1988).

\bibitem{Rietveld} F. Izumi and K. Momma, Solid State Phenom. \textbf{130}, 15 (2007).

\bibitem{Bao} J. K. Bao, J. Y. Liu, C. W. Ma, Z. H. Meng, Z. T. Tang, Y. L. Sun, H. F. Zhai, H. Jiang, H. Bai, C. M. Feng, Z. A. Xu, and G. H. Cao, Phys. Rev. X, \textbf{5}, 011013 (2015).
\bibitem{nickel} F. Ronning, E. D. Bauer, T. Park, N. Kurita, T. Klimczuk, R. Movshovich, A. S. Sefat, D. Mandrus, and J. D. Thompson, Physica C \textbf{469}, 396-403 (2009).
\bibitem{Varma76} C. M. Varma, Rev. Mod. Phys. \textbf{48}, 219-238 (1976).
\bibitem{Hall} S. Nair, S. Wirth, S. Friedemann, F. Steglich, Q. Si, and A. J. Schofield, Advanc. Phys. \textbf{61}, 583-664 (2012).
\bibitem{AHE} N. Nagaosa, J. Sinova, S. Onoda, A. H. MacDonald, and N. P. Ong, Rev. Mod. Phys. \textbf{82}, 1539-1592 (2010).
\bibitem{Lohneysen98} H. von L\"{o}hneysen, A. Neubert, T. Peitrus, A. Schr\"{o}der, O. Stockert, U. Tutsch, M. Loewenhaupt, A. Rosch, and P. W\"{o}lfle, Eur. Phys. J. B \textbf{5}, 447-455 (1998).
%\bibitem{Sumiyama} A. Sumiyama, Y. Oda, H. Nagano, Y. Onuki, K. Shibutani, and Y. Komatsubara, J. Phys. Soc. Jap. \textbf{55}, 1294-1304 (1986).
\bibitem{Pnictide} D. C. Johnston, Advanc. Phys. \textbf{59}, 803-1061 (2010).

\bibitem{CeCu2Si2} H. Q. Yuan, F.M. Grosche, M. Deppe, C. Geibel, G. Sparn, and F. Steglich, Science \textbf{302}, 2104 (2003).
\bibitem{CeCu2Ge2} G. Seyfarth, A.-S. R\"{u}etschi, K. Sengupta, A. Georges, D. Jaccard, S. Watanabe, and K. Miyake, Phys. Rev. B \textbf{85}, 205105 (2012).
\bibitem{CeAu2Si2} Z. Ren, L. V. Pourovskii, G. Giriat, G. Lapertot, A. Georges, and D. Jaccard, Phys. Rev. X \textbf{4}, 031055 (2014).

\bibitem{K-W1} K. Kadowaki and S. B. Woods, Solid State Commun. \textbf{58}, 507-509 (1986).
\bibitem{K-W2} N. Tsujii, H. Kontani, and K. Yoshimura, Phys. Rev. Lett. \textbf{94}, 057201 (2005).

\bibitem{Hertz} J. Hertz, Phys. Rev. B \textbf{14}, 1165-1184 (1974).
\bibitem{Moriya} T. Moriya and T. Takimoto, J. Phys. Soc. Jpn. \textbf{64}, 960 (1995).
\bibitem{Millis} A. J. Millis, Phys. Rev. B \textbf{48}, 7183 (1993).
\bibitem{local} Q. M. Si, S. Rabello, K. Ingersent, and J. U. Smith, Nature, \textbf{413}, 804-808 (2001).
\bibitem{Gegenwart} P. Gegenwart, Q. M. Si, and F. Steglich, Nature Phys. \textbf{4}, 186-197 (2008).
\bibitem{CeNiAsO} Y. K. Luo, L. Pourovskii, S. E. Rowley, Y. K. Li, C. M. feng, A. Georges, J. H. Dai, G. H. Cao, Z. A. Xu, Q. M. Si, and N. P. Ong, Nature Mater. \textbf{13}, 777-781 (2014).
\bibitem{Ir-Yb122} S. Friedemann, T. Westerkamp, M. Brando, N. Oeschler, S. Wirth, P. Gegenwart, C. Krellner, C. Geibel, and F. Steglich, Nature Phys. \textbf{5}, 465¨C469 (2009).
\bibitem{Ge-Yb122} J. Custers, P. Gegenwart, C. Geibel, F. Steglich, P. Coleman, and S. Paschen, Phys. Rev. Lett. \textbf{104}, 186402 (2010).
\bibitem{global} Q. M. Si, Phys. Status Solidi B \textbf{247}, 476-484 (2010).
\bibitem{SiPaschen}
Q. Si and S. Paschen, Phys. Status Solidi B {\bf 250}, 425-438 (2013).


\end{thebibliography}
\end{document}